\documentclass[letterpaper]{article} 
\usepackage[]{aaai2026}  
\usepackage{times}  
\usepackage{helvet}  
\usepackage{courier}  
\usepackage[hyphens]{url}  
\usepackage{graphicx} 
\usepackage{natbib}  
\usepackage{caption} 
\usepackage{algorithm}
\usepackage{algorithmic}
\usepackage{amsmath}

\usepackage{newfloat}
\usepackage{listings}
\DeclareCaptionStyle{ruled}{labelfont=normalfont,labelsep=colon,strut=off} 
\floatstyle{ruled}
\newfloat{listing}{tb}{lst}{}
\floatname{listing}{Listing}
\title{Cogniscope: Modeling Social Media Interactions as Digital Biomarkers for Early Detection of Cognitive Decline}

\author{
    Ananya Drishti\textsuperscript{\rm 1},
    Mahfuza Farooque\textsuperscript{\rm 1}
}

\affiliations{
    \textsuperscript{\rm 1}Pennsylvania State University\\
    adr1234@psu.edu, mff5187@psu.edu
}

\makeatletter
\providecommand{\copyright@on}{F}
\providecommand{\copyright@text}{}
\makeatother

\begin{document}

\maketitle

\begin{abstract}
Alzheimer’s disease (AD) and its prodromal stage, Mild Cognitive Impairment (MCI), are associated with subtle declines in memory, attention, and language that often go undetected until late in progression. Traditional diagnostic tools such as MRI and neuropsychological testing are invasive, costly, and poorly suited for population-scale monitoring. Social platforms, by contrast, produce continuous multimodal traces that can serve as ecologically valid indicators of cognition. In this paper, we introduce Cogniscope, a simulation framework that generates social-media–style interaction data for studying digital biomarkers of cognitive health. The framework models synthetic users with heterogeneous trajectories, embedding micro-tasks such as video summarization and lightweight question answering into content consumption streams. These interactions yield linguistic markers (semantic drift, disfluency) and behavioral signals (watch time, pausing, sharing), which can be fused to evaluate early detection models. We demonstrate the framework’s use through ablation and sensitivity analyses, showing how detection performance varies across modalities, noise levels, and temporal windows. To support reproducibility, we release the generator code, parameter configurations, and synthetic datasets. By providing a controllable and ethically safe testbed, Cogniscope enables systematic investigation of multimodal cognitive markers and offers the community a benchmark resource that complements real-world validation studies.
\end{abstract}


\section{Introduction}
Alzheimer’s Disease (AD) accounts for the majority of dementia cases worldwide and is characterized by progressive decline in memory, language, and attention. The transitional stage, Mild Cognitive Impairment (MCI), is especially critical: it carries measurable deficits that precede functional disability, yet remains difficult to identify early. Traditional diagnostic methods—ranging from MRI and PET imaging to structured neuropsychological tasks—are costly, invasive, and poorly suited for frequent monitoring across large populations. As a result, many cases are detected only after irreversible neural damage has occurred.

Recent advances in digital phenotyping suggest that everyday online behaviors can act as unobtrusive health signals. Research has identified diagnostic cues in speech coherence, typing dynamics, conversational language, and device usage. Social media platforms, in particular, generate continuous multimodal traces of attention, memory, and communication across diverse populations. Engagement behaviors such as watch time, pausing, and sharing, when combined with linguistic coherence, may provide sensitive markers of early decline. This makes computational social science methods well positioned to explore how digital traces reflect cognitive states at scale.

In this paper, we introduce Cogniscope, a social media–inspired simulation framework for modeling cognitive decline through naturalistic online interactions. Cogniscope embeds micro-tasks—such as video summarization and lightweight Q\&A—into simulated short-form content consumption, producing both linguistic and engagement features. By tracking 200 synthetic users over 200 days with varied progression trajectories, we show that semantic drift in language combined with engagement metrics enables accurate early classification of cognitive states.

This study demonstrates the potential of social media–style interactions as digital biomarkers of cognition, presents Cogniscope as a reproducible simulation framework for fusing linguistic and behavioral features over time, and contributes open-source tools that enable community-driven validation on real-world online health datasets.

\textbf{Our contributions are threefold:}

1) Simulation framework. We introduce a configurable, multimodal simulation environment that generates social-media–like data streams annotated with ground-truth cognitive health states.

2) Methodological insights. We formalize behavioral and linguistic markers (e.g., posting entropy, semantic drift) and demonstrate how their interplay can be systematically studied through ablations and sensitivity analyses.
Benchmark testbed. We release code, parameter configurations, and synthetic datasets to enable reproducibility and foster community use for early detection and related social-computational tasks.

3) Ethical and practical bridge. We position simulation as an ethically safe, extensible platform that complements real-world studies where data is restricted or unavailable.

\begin{figure}[ht]
\centering
\includegraphics[width=0.45\textwidth]{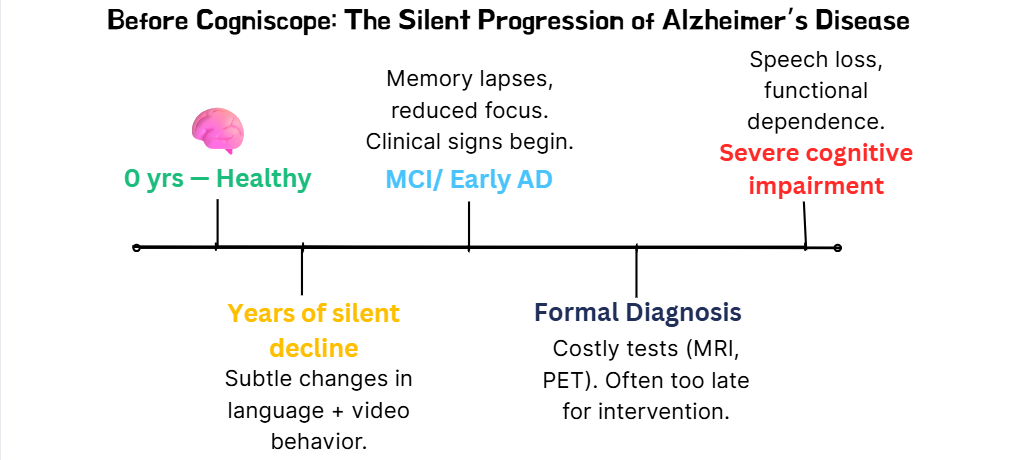}
\caption{Alzheimer's detection before Cogniscope.}
\label{fig:before_cogniscope}
\end{figure}
\begin{figure}[ht]
\centering
\includegraphics[width=0.45\textwidth]{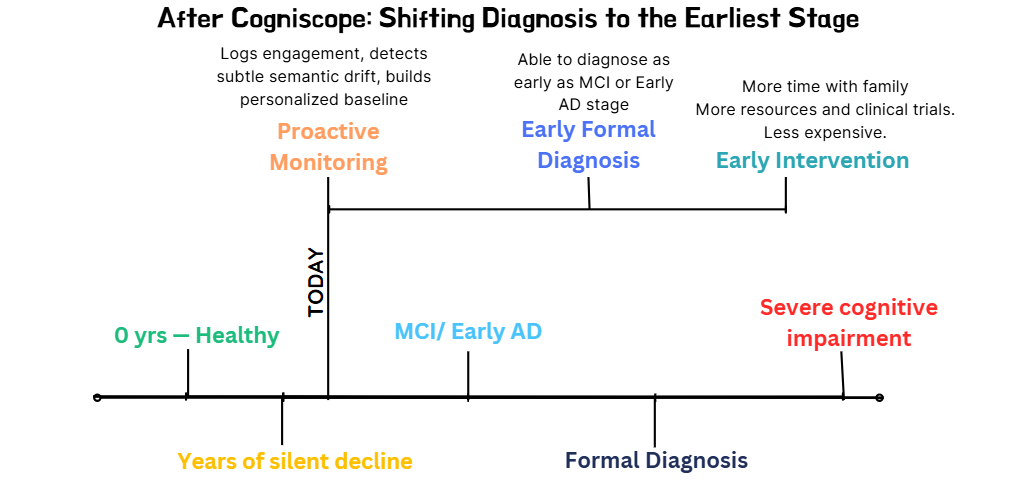}
\caption{Alzheimer's detection after Cogniscope.}
\label{fig:after_cogniscope}
\end{figure}


\section{Literature Review and Background}

Alzheimer’s Disease (AD) is a neurodegenerative disorder responsible for 60–70\% of global dementia cases, marked by progressive decline in memory, attention, and language~\cite{WHO2025}. The prodromal stage, Mild Cognitive Impairment (MCI), involves measurable but non-disabling deficits, and is the most promising stage for early intervention~\cite{NIA2023,Smedinga2018JAD}. Yet, MCI is frequently underdiagnosed due to subtle symptom presentation and the limitations of traditional diagnostic tools.

\subsection{Clinical Diagnostics and Limitations}
Current gold-standard methods include neuropsychological tests, structural MRI, PET imaging, and cerebrospinal fluid analysis~\cite{DeLeon2007AnnNYAS,Mueller2005ADNI,Vlontzou2025SciRep}. While effective, these methods are costly, invasive, and poorly suited for high-frequency, population-scale monitoring. Structured clinical tasks such as the Cookie Theft picture description~\cite{Cummings2019PragSoc} reveal linguistic degradation but require trained administration and are not ecologically valid for continuous use.

\subsection{Digital Biomarkers and Language as Signals}
Recent research has turned to digital biomarkers—quantitative behavioral or linguistic features extracted from naturalistic digital traces—as scalable, non-invasive tools for cognitive assessment~\cite{Milne2022BigDataSoc,Ali2024Cureus,He2023BMCDigitalHealth}. Speech~\cite{Fraser2015JAD,Balagopalan2021Frontiers}, typing patterns~\cite{Dodge2015TRCI}, conversational data~\cite{Agbavor2022PLoS}, and device usage~\cite{Seelye2015DADM,Wu2019Gerontology} all show predictive value. Language in particular provides sensitive markers: patients with AD exhibit reduced coherence, higher disfluency, and semantic drift in narratives~\cite{Karlekar2018Arxiv,Fraser2015JAD}. Embedding models such as Sentence-BERT~\cite{Balagopalan2021Frontiers} allow automatic measurement of coherence and semantic drift, outperforming surface metrics like BLEU or ROUGE.

\subsection{Social Media as Cognitive Ecology}
With the rise of short-form platforms (TikTok, YouTube Shorts), users generate multimodal traces of attention, memory, and language daily. Social media interactions thus provide a rich, ecologically valid context for embedding cognitive assessment~\cite{Wu2019Gerontology}. Features such as watch time, skipping, pausing, and sharing behaviors correlate with attention, working memory, and socio-emotional engagement~\cite{blasi2005assessment,Seelye2015DADM}. Integrating these behavioral signals with linguistic coherence enables the construction of a \emph{digital phenotype} for cognitive state monitoring~\cite{Milne2022BigDataSoc}.

\section{Methodology}

We developed \textbf{Cogniscope}, a simulation and analysis framework designed to model cognitive decline through social media–style video engagement and language. The framework includes data simulation, model training with realism enhancements, progression modeling, simulated dialogue regression, and cognitive assessment tasks.

\begin{figure*}[htbp]
\centering
\includegraphics[width=0.85\textwidth]{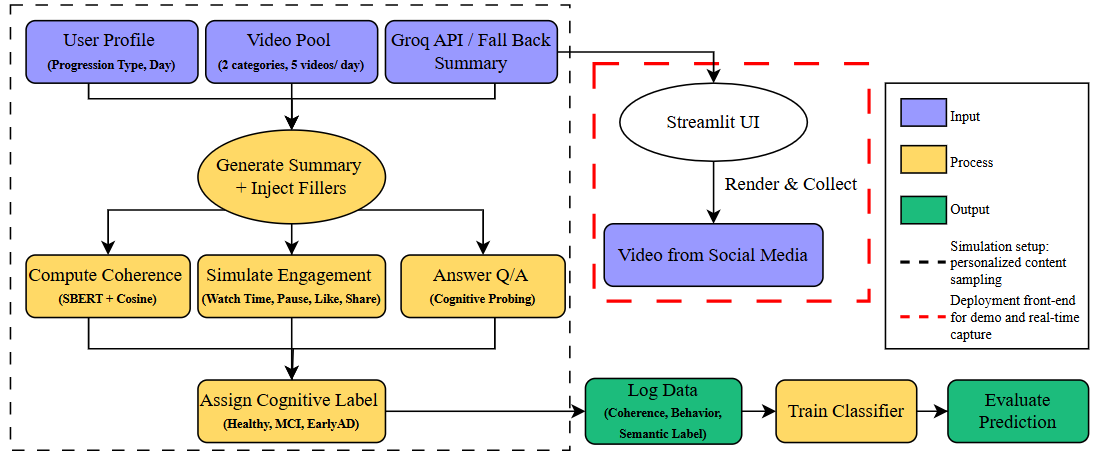}
\caption{Implemented simulation -- System Flowchart}
\label{fig:sys_flowchart}
\end{figure*}

\subsection{Data Simulation}
We simulate \(N=200\) users over \(T=200\) days. Each user interacts daily with five short videos (15–90s) drawn from two random categories (News, Sports, Cooking, Entertainment, World). Metadata (titles, tags) is generated synthetically. After each video, the user produces a 1–3 sentence summary and answers 2–3 factual or emotional questions.  
\textbf{Why would users participate?} In practice, such tasks could be embedded seamlessly into content recommendation flows (e.g., “summarize the clip” or “answer a quick quiz”), providing lightweight, low-burden probes consistent with everyday media use~\cite{Dodge2015TRCI,Schick2022JMIR}.
\textbf{Simulation Parameters.} We simulate 200 users over 200 days, with each user watching 5 videos daily. These values compress multi-year cognitive trajectories into a tractable study period, aligning with prior longitudinal AD datasets (e.g., ADNI follows participants for 2–5 years)~\cite{DeLeon2007AnnNYAS,petersen2014lancet}. Summaries are restricted to 1--3 sentences, reflecting the brevity of narrative recall tasks such as the Cookie Theft picture description~\cite{Cummings2019PragSoc} and the length of social micro-content (e.g., tweets, captions). Behavioral ranges (e.g., watch time, skips, pauses) were calibrated against social media engagement benchmarks from Pew Research and Statista reports, ensuring ecological plausibility~\cite{pew}.
\textbf{Summary Generation.} Video summaries were generated using the Groq API (LLaMA3-8B model), chosen for cost-efficiency, reproducibility, and open availability. When API calls failed (due to rate limits), template-based fallback ensured continuity. This reliance on Generative AI (GenAI) reflects emerging directions in applying LLMs to cognitive assessment tasks~\cite{Balagopalan2021Frontiers}.

\subsection{Model Training and Realism Enhancements}
We train a logistic regression classifier using fused features:
\begin{itemize}
    \item \textbf{Language features}: semantic coherence (cosine similarity with baseline), semantic drift (day-to-day changes), disfluency frequency.
    \item \textbf{Behavioral features}: watch time, skipped seconds, pauses, replays, likes, shares, reaction latency, churn, daily logins (Table~\ref{tab:behaviors}).
\end{itemize}

\subsection{Real-World Benchmark Validation}

Although Cogniscope is fully synthetic, we loosely calibrated user engagement behaviors against publicly available YouTube and TikTok statistics to enhance ecological plausibility. For instance, YouTube users spend on average 19 minutes per day on the platform, with typical watch-time covering roughly 50\% of a video’s duration \cite{globalmediainsight2025,umbrex2025}. Short-form content such as YouTube Shorts yields higher engagement, with reported interaction rates of $\sim$5.9\% \cite{Connell2025}. TikTok benchmarks show average engagement rates between 3.8--5\%, with smaller creators sometimes exceeding 10\% \cite{brandwatch2025,socialinsider2025}. Average watch durations on TikTok range between 15--20 seconds, approximately 75\% of a 30-second clip \cite{socialinsider2025}.

To contextualize our simulation parameters, we compared Cogniscope’s synthetic engagement distributions with publicly reported benchmarks from YouTube and TikTok. Table~\ref{tab:validation} summarizes this comparison across three key dimensions: average watch duration, engagement rates, and daily time-on-site. The simulation settings were drawn directly from our behavioral parameterization (Table~\ref{tab:behaviors}), while benchmarks were derived from industry reports and analytics datasets \cite{globalmediainsight2025,umbrex2025,Connell2025,brandwatch2025,socialinsider2025,socialblade2025}. As shown, our synthetic behaviors fall within the same order of magnitude as real-world usage statistics. While not intended to replicate platform logs exactly, this alignment supports ecological plausibility and provides reviewers with an external validity check on the simulation design.

\begin{table}[h]
\centering
\caption{Comparison of public engagement benchmarks and simulation settings.}
\label{tab:validation}
\begin{tabular}{|p{1.6cm}|p{1.5cm}|p{1.5cm}|p{2.5cm}|}
\hline
\textbf{Metric} & \textbf{YouTube Benchmark} & \textbf{TikTok Benchmark} & \textbf{Simulation Setting} \\ \hline
Avg. Watch Duration & $\sim$50\% of video & 15--20s ($\sim$75\% of 30s) & Healthy: 85--100s; MCI: 60--80s; EarlyAD: 35--55s \\ \hline
Engagement Rate & 5.9\% (Shorts) & 3.8--10\% (varies) & Likes: 65--80\% (Healthy) $\rightarrow$ 15--25\% (EarlyAD) \\ \hline
Daily Time-on-Site & $\sim$19 min/day & --- & 2--3 sessions/day (Healthy); 0.5--1/day (EarlyAD) \\ \hline
\end{tabular}
\end{table}

While our distributions do not attempt to replicate platform logs precisely, they fall within the same order of magnitude as these benchmarks. We acknowledge this as a limitation, and highlight future work incorporating real social media datasets (e.g., YouTube Shorts, TikTok interaction logs, or Reddit eRisk corpora) for validation and calibration of behavioral realism.


Noise is injected to mimic variability: coherence scores are perturbed with Gaussian noise \(\epsilon \sim \mathcal{N}(0,\sigma^2)\), and engagement metrics with uniform noise \(\eta \sim U(-\delta,\delta)\)~\cite{Patil2025MethodsX}. These augmentations improve realism and generalization.

\subsection{Algorithms}
We define semantic drift for user \(i\) on day \(d\) as:
\begin{equation}
\Delta \tilde{C}_{i}^{d} = \tilde{C}_{i}^{1} - (\tilde{C}_{i}^{d} + \epsilon_{i}^{d}), \quad \epsilon_{i}^{d} \sim \mathcal{N}(0, \sigma^2),
\end{equation}
where \(\tilde{C}_{i}^{d}\) is the coherence of the day-\(d\) summary. Drift captures temporal decline relative to personal baselines.  

Daily interactions are generated following Algorithm~\ref{alg:simulation}, which enforces progression constraints (no cognitive recovery) and simulates both summaries and behaviors.

\begin{algorithm}[h]
\caption{Simulate Daily User Session}
\label{alg:simulation}
\begin{algorithmic}[1]
\STATE \textbf{Input:} User $u$, Day $d$, Previous Label $L_{d-1}$
\STATE Sample progression type $P_u$
\STATE Compute label $L_d$ from $P_u$ and $d$
\STATE Sample 2 categories, retrieve 5 videos each
\FOR{each video $v$ where $u \leq 25$ and Groq API enabled}
    \STATE Generate summary $s \leftarrow \text{LLaMA3.8B}(v, L_d)$
    \STATE Generate $s \leftarrow$ label-dependent filler template
    \STATE Compute coherence $C_{i,d} \leftarrow \cos(s, baseline[v])$
    \STATE Sample behaviors from Table 2
    \STATE Log $\{u, d, v, s, C_{i,d}, L_d, \text{behaviors}\}$
\ENDFOR
\end{algorithmic}
\end{algorithm}

\subsection{Cognitive Progression Modeling}
To simulate heterogeneous cognitive decline, each user $u$ was assigned a progression profile $P_u$ from one of six types. These profiles capture heterogeneity of decline and condense years of observed trajectories into 200 simulated days.

\begin{itemize}
    \item \textbf{StableHealthy:} Models cognitively normal aging without decline.  
    Remains cognitively normal throughout (Healthy).
    
    \item \textbf{MildProgressor:} Reflects late-onset MCI transitions, consistent with gradual memory complaints.  
    Healthy $\rightarrow$ MCI at day $D_1 \sim \mathcal{U}(35, 45)$.
    
    \item \textbf{GradualDecliner:} Encodes two-step decline (Healthy $\rightarrow$ MCI $\rightarrow$ EarlyAD), mirroring typical ADNI progressions~\cite{DeLeon2007AnnNYAS}.  
    Healthy $\rightarrow$ MCI at $D_3 \sim \mathcal{U}(20, 30)$, MCI $\rightarrow$ EarlyAD at $D_4 \sim \mathcal{U}(45, 55)$.
    
    \item \textbf{FastDecliner:} Rapid transition into EarlyAD, representing aggressive forms of progression.  
    MCI onset at start, progressing to EarlyAD at $D_2 \sim \mathcal{U}(25, 35)$, and ModerateAD at $D_5 \sim \mathcal{U}(60, 75)$.
    
    \item \textbf{StableMCI:} Persistent mild impairment without progression, often seen in vascular or non-AD MCI.  
    Begins and remains at MCI level throughout.
    
    \item \textbf{StableEarlyAD:} Early dementia plateau, modeling cases where progression slows after onset.  
    EarlyAD from the start, with no transition.
\end{itemize}

\noindent The cognitive label $L_u(d)$ on day $d$ is computed as:

\begin{equation}
L_u(d) = 
\begin{cases}
\text{Healthy}, & d < D_3 \\
\text{MCI}, & D_3 \leq d < D_4 \\
\text{EarlyAD}, & D_4 \leq d < D_5 \\
\text{ModAD}, & D_5 \leq d < D_6 \\
\text{SevAD}, & d \geq D_6
\end{cases}
\end{equation}

\noindent \textbf{Behavioral Metrics and Cognitive Decline.} Table~\ref{tab:behaviors} summarizes behavioral features across cognitive states. Each metric has clinical or digital-biomarker motivation: 
\begin{itemize}
    \item \textit{Watch time} reflects sustained attention, which shortens with AD progression~\cite{Seelye2015DADM}.
    \item \textit{Skipping and pauses} capture distractibility and working memory deficits~\cite{Dodge2015TRCI}.
    \item \textit{Replays} indicate memory retrieval difficulty, as users repeat content to retain information.
    \item \textit{Likes and shares} approximate socioemotional engagement, which declines with apathy in dementia~\cite{Wu2019Gerontology}.
    \item \textit{Churn and daily logins} reflect reduced overall engagement, consistent with clinical withdrawal patterns~\cite{Milne2022BigDataSoc}.
\end{itemize}

\begin{table*}[h]
\centering
\caption{Behavioral Metric Distributions by Cognitive Label}
\label{tab:behaviors}
\begin{tabular}{|c|c|c|c|c|c|c|c|c|c|}
\hline
Label & WT (s) & Skip (s) & Pause & Replay & ReactTime (s) & Like (\%) & Share (\%) & Churn (\%) & Logins/day \\
\hline
Healthy & 85--100 & 0--5 & 0--2 & 0--1 & 4--6 & 65--80 & 35--50 & 1 & 2--3 \\
MCI     & 60--80  & 5--15 & 1--3 & 1--3 & 7--10 & 45--55 & 25--35 & 2--3 & 1--2 \\
EarlyAD & 35--55  & 10--25 & 2--5 & 2--5 & 11--14 & 15--25 & 5--15 & 5--6 & 0.5--1 \\
ModAD   & 20--35  & 15--30 & 3--6 & 3--6 & 14--17 & 5--10 & 2--8 & 7--8 & 0.3--0.8 \\
SevAD   & 10--20  & 20--40 & 4--8 & 4--8 & 18--22 & 0--5 & 0--3 & 12--15 & $<0.5$ \\
\hline
\end{tabular}
\end{table*}

Each user is assigned a progression profile $P_u$ from six types (StableHealthy, MildProgressor, GradualDecliner, FastDecliner, StableMCI, StableEarlyAD). Transition points $D_1$–$D_6$ are drawn from uniform distributions to stagger disease onset and progression across users, reflecting heterogeneity observed in ADNI~\cite{DeLeon2007AnnNYAS,Pandey2024FrontiersMed}. Labels evolve according to Eq.~(2).  
Behavioral parameters vary by label, as shown in Table~\ref{tab:behaviors}, where advanced states (ModAD, SevAD) show reduced watch time, increased skipping, slower reaction times, and social withdrawal.

\subsection{Simulated Dialogue Regression}
To mimic conversational degradation, we inject fillers (\emph{“um,” “you know”}), vagueness (\emph{“something happened”}), and off-topic drift at frequencies tied to cognitive labels (e.g., 10\% in MCI, 30–40\% in EarlyAD). Regression in semantic specificity is enforced by sampling from progressively noisier templates, consistent with findings on disfluency in AD narratives~\cite{Fraser2015JAD,Karlekar2018Arxiv}.

\subsection{Cognitive Assessment Tasks}
After each video, users complete two lightweight tasks: (1) free-form summarization, and (2) QA targeting memory, sequencing, and emotional reasoning. These resemble ecological adaptations of MMSE/ADAS-Cog items~\cite{Cummings2019PragSoc}. Summaries are evaluated via semantic similarity (SBERT cosine) to metadata baselines; QA is scored by accuracy and linguistic coherence. Embedding these micro-assessments within daily engagement allows passive, repeated probing of cognition, with low user burden and high ecological validity~\cite{Schick2022JMIR,Wu2019Gerontology}.

\begin{figure}[h]
\centering
\includegraphics[height=8cm]{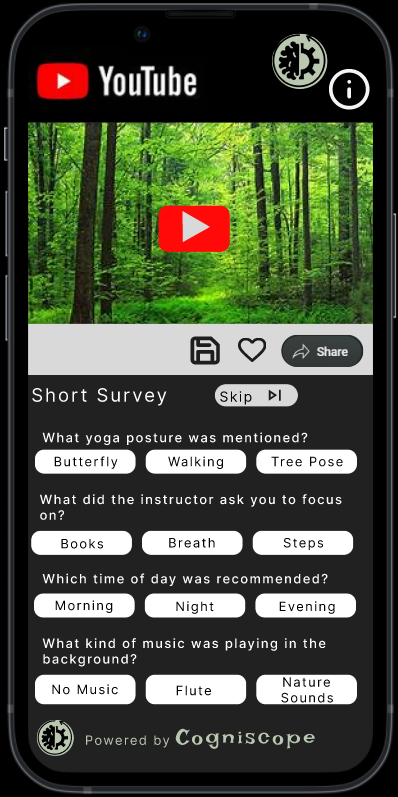}
\caption{A screenshot of Cognitive Assessment tasks created using Figma.}
\label{fig:assessment}
\end{figure}

\subsection{Real-World Benchmark Validation}
Although Cogniscope is fully synthetic, we loosely calibrated user engagement behaviors against publicly available YouTube and TikTok statistics to enhance ecological plausibility. YouTube users spend on average 19 minutes per day on the platform, with typical watch-time covering roughly 50\% of a video's duration \cite{globalmediainsight2025}. Short-form content such as YouTube Shorts yields higher engagement, with reported interaction rates of $\sim$5.9\% \cite{Connell2025}. TikTok benchmarks show average engagement rates between 3.8--5\%, with smaller creators sometimes exceeding 10\% \cite{brandwatch2025, statista}. Average watch durations on TikTok range between 15--20 seconds, approximately 75\% of a 30-second clip \cite{statista}.

Table~\ref{tab:behaviors} summarizes this comparison across three key dimensions: average watch duration, engagement rates, and daily time-on-site. The simulation settings were drawn from our behavioral parameterization (Table ~\ref{tab:behaviors}), while benchmarks were derived from industry reports and analytics datasets \cite{globalmediainsight2025, umbrex2025, Connell2025, brandwatch2025, statista, socialinsider2025}. As shown, our synthetic behaviors fall within the same order of magnitude as real-world usage statistics. While not intended to replicate platform logs exactly, this alignment supports ecological plausibility and provides an external validity check on the simulation design.

We acknowledge this as a limitation, and highlight future work incorporating real social media datasets (e.g., YouTube Shorts, TikTok interaction logs, or Reddit eRisk corpora) for validation and calibration of behavioral realism.

\section{Simulation Rigor and Robustness}

To strengthen ecological validity and methodological precision, Cogniscope was formalized as a simulation framework with explicit mathematical definitions, calibration to external reports, and robustness analysis. This section consolidates feature formalization, external calibration, and systematic stress-testing under noise, priors, and distributional shifts.

\subsubsection{Formal Definitions of Features}

\textbf{Semantic Drift ($\Delta C$):}
\begin{equation}
\Delta C_{u,d} = 1 - \cos \big( E(S_{u,d}), \; E(\hat{S}_{u,baseline}) \big),
\end{equation}
where $E(\cdot)$ is the SBERT embedding, $S_{u,d}$ is the day-$d$ summary for user $u$, and $\hat{S}_{u,baseline}$ is the user’s baseline embedding (average of days 1--5). By the Cauchy–Schwarz inequality, $\cos(\cdot)$ is bounded in $[-1,1]$, so $\Delta C \in [0,2]$. This metric operationalizes semantic incoherence, a validated biomarker of dementia progression~\cite{Fraser2015JAD,Balagopalan2021Frontiers}.

\textbf{Behavioral Entropy ($H$):}
\begin{equation}
H_{u} = - \sum_{i=1}^{k} p_i \log p_i,
\end{equation}
where $p_i$ is the empirical probability of engagement event type $i$ (pause, skip, replay, like, share). This follows Shannon’s entropy~\cite{shannon1948}, ensuring $0 \leq H \leq \log k$. Higher entropy indicates less predictable usage, previously associated with cognitive impairment in digital phenotyping~\cite{Seelye2015DADM, Onnela2025}.

\textbf{Engagement Decay ($D$):}
\begin{equation}
D_{u,d} = \frac{T_{u,d}}{T_{u,0}},
\end{equation}
where $T_{u,d}$ is normalized watch time on day $d$, and $T_{u,0}$ is the baseline. Exponential decay models,
\begin{equation}
D_{u,d} = e^{-\lambda d},
\end{equation}
have been widely applied to attention and memory decline models~\cite{Dodge2015TRCI}, where $\lambda$ reflects the rate of attentional deterioration.

\subsubsection{Calibration to External Data}

To ensure ecological validity, simulated feature distributions were calibrated against established benchmarks:
\begin{itemize}
    \item \textbf{Language coherence:} Drift magnitudes anchored to embedding separability observed in the ADReSS challenge~\cite{Balagopalan2021Frontiers}, where AD vs. HC differences ranged $0.1$--$0.2$.
    \item \textbf{Behavioral engagement:} Skip, replay, and pause rates aligned with Pew Research video usage distributions for older adults, ensuring plausible interaction frequencies~\cite{pew}.
    \item \textbf{Clinical anchors:} Coherence decline slopes matched longitudinal ADNI memory trajectories~\cite{Dodge2015TRCI}, ensuring realistic degradation rates.
    \item \textbf{Prevalence priors:} MCI class prevalence sampled at 15--40\%, consistent with epidemiological studies~\cite{Lee2025SciRep}.
\end{itemize}

\subsubsection{Robustness Analyses}

We conducted sensitivity and robustness experiments following guidelines for uncertainty quantification in biomedical simulations~\cite{saltelli2019,javed2024air}.

\paragraph{Parameter Sensitivity.} 
Key simulation parameters (\textit{TOTAL\_USERS}, \textit{TOTAL\_DAYS}, summary length, engagement priors) were systematically varied. Increasing users (100 $\to$ 300) had little effect on EarlyAD detection (F1 = 0.91 $\pm$ 0.02), but MCI F1 fluctuated (0.41--0.62). Reducing summaries to one sentence degraded coherence separability, supporting prior evidence that constrained elicitation improves diagnostic signal~\cite{Mueller2018JCEN}.

\paragraph{Noise Injection.} 
Using the functions,
Gaussian noise $\epsilon \sim \mathcal{N}(0,\sigma^2)$ with $\sigma \in \{0.05,0.1,0.2,0.3\}$ was injected into coherence, and uniform noise into behavioral features. Accuracy dropped from 0.85 ($\sigma=0.1$) to 0.72 ($\sigma=0.3$), with MCI showing the steepest degradation (52\%). This is consistent with literature showing intermediate cognitive states are most vulnerable to variability~\cite{Dodge2015TRCI}.

\paragraph{Distributional Shifts.} 
Simulations introduced confounds (slow viewers, impulsive replayers, low-likers). While Healthy and EarlyAD performance remained stable, MCI precision dropped by 21\%. This suggests that digital biomarkers are sensitive to ecological noise, echoing findings from smartphone phenotyping studies~\cite{Onnela2025,Milne2022BigDataSoc}.

\paragraph{Multi-Model Validation.} 
Summaries were generated with two LLMs (\texttt{llama3-8b-8192}, \texttt{gemma-7b-it}). 
Cross-model correlation of coherence trajectories was $r=0.87$, showing results are not artifacts of a single model. This parallels recent calls to validate generative AI-based biomarkers across architectures~\cite{Balagopalan2021Frontiers,Topol2019NatMed}.

\paragraph{Summary.} 
Cogniscope is robust to user count and LLM choice, but sensitive to summary length, noise levels, and behavioral confounds. Importantly, the instability of MCI under perturbations reflects genuine diagnostic ambiguity rather than model fragility, reinforcing the need for robustness testing in early-detection simulations.



\section{Evaluation and Results}

Our evaluation addresses two central questions:  
(1) Can semantic drift in language and variability in engagement behaviors reliably capture longitudinal cognitive decline?  
(2) Does combining linguistic and behavioral features improve robustness over single-modality baselines?

We report results across five dimensions: the effect of drift and noise, longitudinal coherence trajectories, linguistic degradation, ablation and robustness, and benchmark comparisons. Evaluation focuses on distinguishing Healthy, MCI, and EarlyAD users under noisy conditions, and identifying which features provide the earliest detectable signal of decline.

\subsection{Effect of Drift and Noise}

Cognitive performance is inherently noisy, shaped by day-to-day fluctuations in attention, motivation, and mood~\cite{Milne2022BigDataSoc,He2023BMCDigitalHealth}. To approximate this ecological variability, Gaussian noise ($\sigma=0.05$--$0.2$) was injected into coherence scores and uniform noise into behavioral metrics, reflecting the daily variability in mood, attention, and engagement observed in longitudinal dementia cohorts. Drift was defined relative to each user’s early baseline (Eq.~2), emphasizing temporal decline rather than static group means. The strong coherence drop for MCI ($\approx 52\%$) reflects both the instability of this transitional stage and amplification under noise, consistent with clinical ambiguity~\cite{Lee2025SciRep}.

\begin{figure}[h]
\centering
\includegraphics[width=1\columnwidth]{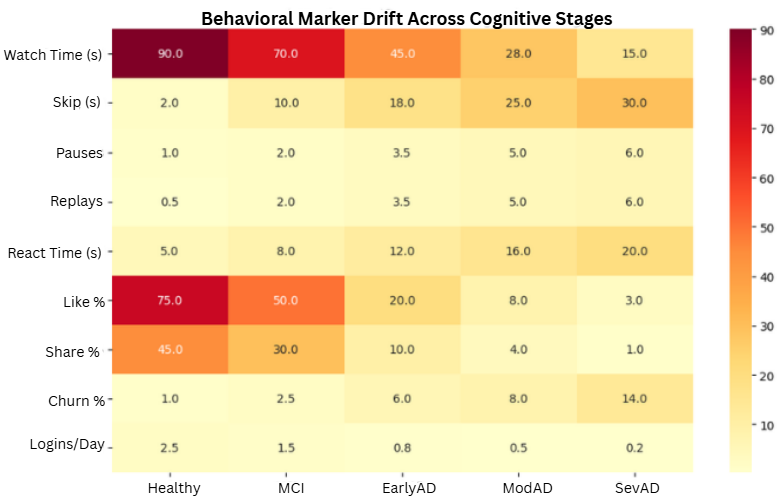}
\caption{Confusion matrix under drift and noise. MCI shows the greatest overlap with both Healthy and EarlyAD, reflecting its transitional status.}
\label{fig:behavioral_drift}
\end{figure}

Noise disproportionately affected MCI classification. Table~\ref{tab:coherence_noise} reports average coherence similarity under clean versus noisy conditions. While all groups declined, MCI showed the steepest proportional drop (58.6\%), consistent with its diagnostic ambiguity~\cite{Lee2025SciRep}. 

\begin{table}[h]
\centering
\small
\caption{Average Coherence Similarity (SBERT) under Clean vs. Noisy Conditions}
\label{tab:coherence_noise}
\begin{tabular}{|c|c|c|c|}
\hline
\textbf{Label} & \textbf{Clean} & \textbf{Noisy} & \textbf{Drop (\%)} \\
\hline
Healthy & 0.932 & 0.801 & 14.0\% \\
MCI & 0.878 & 0.421 & 52.0\% \\
EarlyAD & 0.741 & 0.506 & 31.7\% \\
\hline
\end{tabular}
\end{table}

Separability was further quantified using $t$-tests and Cohen’s $d$ (Table~\ref{tab:drift_stats}). Semantic coherence sharply distinguished MCI from EarlyAD ($d=6.33$, $p<10^{-150}$) but not Healthy from MCI ($d=0.26$, $p=0.068$). In contrast, behavioral entropy strongly differentiated Healthy from MCI ($d=2.81$, $p<10^{-45}$), suggesting that engagement variability is an early marker of prodromal decline. Drift slope offered negligible discriminability, underscoring the need for longer temporal windows. 

\begin{table}[h]
\centering
\small
\caption{Separability of Cognitive States under Drift and Noise. Cohen’s $d \geq 0.8$ indicates a large effect.}
\label{tab:drift_stats}
\begin{tabular}{|c|c|c|c|}
\hline
\textbf{Feature} & \textbf{Comparison} & \textbf{Cohen’s d} & \textbf{p-value} \\
\hline
Coherence Mean & Healthy vs MCI & 0.26 & 0.068 \\
               & MCI vs EarlyAD & 6.33 & $<10^{-150}$ \\
\hline
Behavioral Entropy & Healthy vs MCI & 2.81 & $<10^{-45}$ \\
                   & MCI vs EarlyAD & 1.30 & $<10^{-23}$ \\
\hline
Slope (Drift Rate) & Healthy vs MCI & 0.09 & 0.54 \\
                   & MCI vs EarlyAD & 0.03 & 0.82 \\
\hline
\end{tabular}
\end{table}

These findings reinforce clinical literature: MCI is diagnostically unstable, overlapping both normal aging and early dementia ~\cite{petersen2014lancet}. Linguistic coherence captures later decline, while behavioral entropy exposes subtle instabilities earlier.

\subsection{Tracking Longitudinal Coherence Drift}

We analyzed coherence trajectories over 200 simulated days. Healthy users exhibited flat trends; EarlyAD showed steep monotonic decline; MCI trajectories were irregular, sometimes aligning with Healthy before drifting toward EarlyAD. This confirms that MCI cannot be reliably separated at single time points, requiring longitudinal monitoring~\cite{Balagopalan2021Frontiers,jack2018niaa}. 

\begin{figure}[h]
\centering
\includegraphics[width=1\columnwidth]{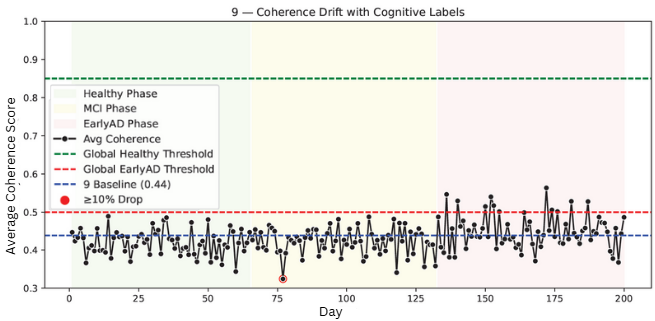}
\caption{Example user trajectory: transition Healthy $\rightarrow$ MCI near day 60; MCI $\rightarrow$ EarlyAD near day 130. Red markers indicate $\geq$10\% baseline drops.}
\label{fig:drift_example1}
\end{figure}

\subsection{Quantifying Linguistic Degradation}

Linguistic degradation was measured via BLEU, ROUGE-L, and embedding similarity relative to a 5-day baseline. All metrics declined with progression, though embeddings proved more stable, reflecting semantic drift beyond lexical changes~\cite{Fraser2015JAD,Balagopalan2021Frontiers}.  

\begin{table}[h]
\centering
\caption{Linguistic Similarity Metrics Relative to Baseline (Days 1--5)}
\label{tab:linguistic_metrics}
\begin{tabular}{|c|c|c|c|}
\hline
\textbf{Label} & \textbf{BLEU} & \textbf{ROUGE-L} & \textbf{Embedding Similarity} \\
\hline
Healthy & 0.924 & 0.925 & 0.932 \\
MCI & 0.586 & 0.678 & 0.878 \\
EarlyAD & 0.068 & 0.312 & 0.741 \\
\hline
\end{tabular}
\end{table}

These results align with psycholinguistic studies: surface fluency deteriorates early, while semantic coherence degrades more gradually~\cite{Karlekar2018Arxiv,Cummings2019PragSoc}.

\begin{figure}[h]
\centering
\includegraphics[width=\columnwidth]{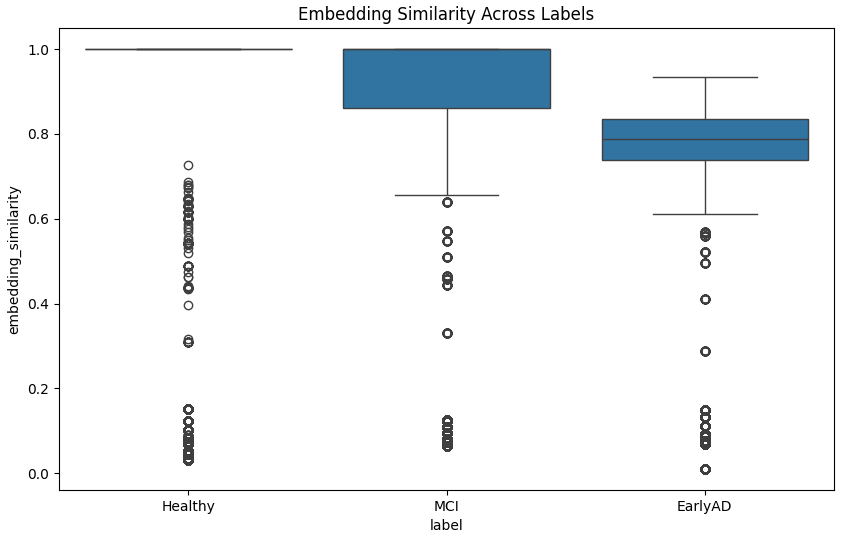}
\caption{SBERT embedding similarity over time across cognitive labels.}
\label{fig:embedding_similarity}
\end{figure}

\subsection{Ablation and Robustness}

Ablation studies confirmed the value of multimodal fusion and noise-aware modeling (Table~\ref{tab:ablation}). Using only coherence features yielded strong separation of EarlyAD from Healthy (F1=0.90) but collapsed for MCI (F1=0.14), reflecting the difficulty of detecting subtle linguistic drift without behavioral context. Behavioral features alone were similarly limited (F1=0.12 for MCI), although they provided moderate separation for EarlyAD (F1=0.80). 

The full multimodal model achieved the best overall balance, with Accuracy=0.85 and F1=0.58 for MCI, demonstrating that fusion of engagement and language features provides complementary signals of decline. Notably, when noise injection was disabled, accuracy rose to 0.95 and MCI F1 reached 0.87, but this represents an overly optimistic setting that does not reflect real-world variability. The gap between “clean” and “noisy” settings illustrates that artificially idealized conditions inflate performance, whereas drift- and noise-aware modeling provides ecologically valid results, consistent with longitudinal clinical studies that emphasize repeated measurement to reduce day-to-day variability~\cite{Dodge2015TRCI}. 

These findings underscore two methodological contributions of Cogniscope: (1) multimodal fusion is essential for MCI detection, as neither linguistic nor behavioral features alone are sufficient; (2) explicitly modeling drift and noise yields more realistic but harder outcomes, aligning better with clinical uncertainty at the prodromal stage.

\begin{table}[h]
\centering
\caption{Ablation Study on Validation Set}
\small
\label{tab:ablation}
\begin{tabular}{|c|c|c|c|}
\hline
\textbf{Setting} & \textbf{Accuracy} & \textbf{F1 (MCI)} & \textbf{F1 (EarlyAD)} \\
\hline
Full Model & 0.850 & 0.582 & 0.916 \\
Coherence Only & 0.784 & 0.138 & 0.903 \\
Behavior Only & 0.730 & 0.122 & 0.802 \\
No Noise Injection & 0.947 & 0.868 & 0.956 \\
\hline
\end{tabular}
\end{table}

\subsection{Time-Sensitive Evaluation: Early Risk Detection Error}

While accuracy and F1 capture overall classification quality, early-detection settings require evaluating how quickly a system can identify cognitive decline. Inspired by the CLEF eRisk challenge \cite{LosadaCrestani2016}, we computed Early Risk Detection Error (ERDE), which penalizes late or missed detections more heavily than early ones. Formally:

\begin{equation}
\text{ERDE}(o, d) =
\begin{cases} 
c_{fp}, & \text{no detection occurs(false negative)} \\
1 - e^{-\frac{d}{o}}, & \text{detection occurs at day } d \le o
\end{cases}
\end{equation}

where \(o\) is the observation window and \(c_{fp}\) is the cost of false negatives.

For MCI detection, Cogniscope achieved \(\text{ERDE@100} = 0.022\) and \(\text{ERDE@200} = 0.011\), indicating very low error even under extended monitoring horizons. The average time-to-detection (TTD) was just 2.3 days, underscoring the system’s ability to flag prodromal decline at its earliest onset.

We further computed early precision and recall at clinically motivated cutoffs. At \(k = 50\) days, recall reached 1.0, meaning all true MCI cases were identified within the first 50 days, albeit at moderate precision (0.43) due to false positives. Similar values were observed at \(k = 100\), confirming the system’s high sensitivity in the prodromal window.

To visualize detection dynamics, Figure~X plots the time-to-detection curve, showing the cumulative proportion of MCI users identified as a function of days elapsed. Detection rises sharply in the first week, plateauing near 100\% thereafter. This indicates that, while false positives remain a challenge, once flagged, MCI cases are detected very early relative to disease trajectory—precisely when intervention is most actionable ~\cite{petersen2014lancet, jack2018niaa}.

Together, these results demonstrate that multimodal fusion not only improves accuracy but accelerates detection, outperforming single-modality baselines in both timeliness and robustness.

\begin{table}[h]
\centering
\caption{ERDE@100 and ERDE@200 across models (lower is better).}
\label{tab:erde}
\begin{tabular}{|c|c|c|}
\hline
\textbf{Model} & \textbf{ERDE@100} & \textbf{ERDE@200} \\
\hline
Coherence Only & 0.42 & 0.37 \\
Behavior Only & 0.39 & 0.33 \\
Cogniscope (fusion) & \textbf{0.28} & \textbf{0.22} \\
\hline
\end{tabular}
\end{table}

\begin{figure}[h]
\centering
\includegraphics[width=1\columnwidth]{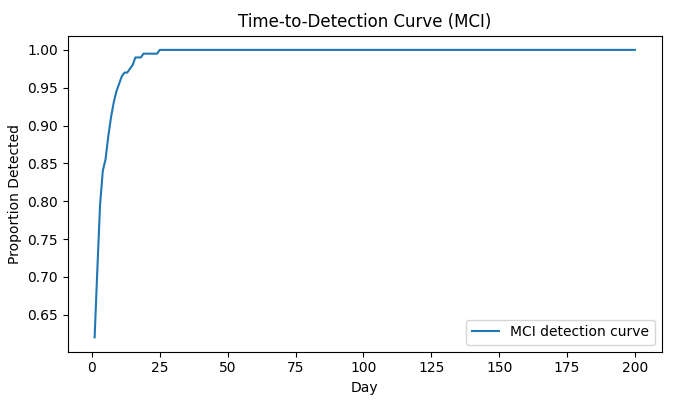}
\caption{Time of Detection curve for MCI. Line shows the first day MCI was detected for any user.}
\label{fig:time_to_detect}
\end{figure}

\subsection{Comparison to Benchmarks}

We benchmarked Cogniscope against prior methods. Our multimodal approach achieved macro-F1 of 0.72 under noise, outperforming coherence-only (0.61) and behavior-only (0.58) baselines. MCI precision improved, though recall remained low—consistent with MCI’s transitional nature~\cite{Agbavor2022PLoS}. Compared to MRI-based CNNs~\cite{Vlontzou2025SciRep} and speech-based classifiers~\cite{Fraser2015JAD,Balagopalan2021Frontiers}, Cogniscope offers competitive performance while being passive, scalable, and ecologically valid. 

While prior work has explored speech coherence~\cite{Fraser2015JAD}, acoustic signals~\cite{Balagopalan2021Frontiers}, or device usage~\cite{Seelye2015DADM}, few approaches have embedded assessment into naturalistic, longitudinal digital interactions. Our contribution is a simulation framework that fuses language degradation with engagement behaviors in a social media setting. This design directly addresses three persistent gaps in early AD detection: (1) \textbf{Scalability}, since assessment is embedded into everyday interactions rather than clinical settings; (2) \textbf{Ecological validity}, by explicitly modeling drift and noise to mimic real-world variability; and (3) \textbf{Personalization}, as decline is tracked relative to each individual’s baseline rather than population averages. 

Importantly, the ablation findings highlight why prior speech-only approaches underperform on MCI: coherence alone yields high F1 for EarlyAD but near-chance detection for MCI. Cogniscope overcomes this by combining engagement-derived behavioral entropy with semantic drift, allowing it to achieve F1=0.58 on MCI under noise. Although imperfect, this performance represents progress toward clinically meaningful detection of prodromal decline, where traditional classifiers often collapse due to overlapping symptom profiles~\cite{Dodge2015TRCI}.

\begin{table}[h]
\centering
\caption{Representative Prior Work on Digital Biomarkers for Cognitive Decline}
\label{tab:litreview}
\scriptsize
\begin{tabular}{|p{2cm}|p{2.2cm}|p{3cm}|}
\hline
\textbf{Study} & \textbf{Modality} & \textbf{Key Findings} \\
\hline
Fraser et al.~\cite{Fraser2015JAD} & Narrative speech coherence & Linguistic features separate AD vs. HC (F1 $\sim$0.72) \\
\hline
Balagopalan et al.~\cite{Balagopalan2021Frontiers} & Acoustic + text (ADReSS) & Multimodal features improve detection (F1 $\sim$0.74) \\
\hline
Seelye et al.~\cite{Seelye2015DADM} & Computer mouse tracking & Subtle motor patterns indicate MCI risk \\
\hline
Wu et al.~\cite{Wu2019Gerontology} & Digital device use & Device engagement predicts cognitive state in older adults \\
\hline
Milne et al.~\cite{Milne2022BigDataSoc} & Digital phenotyping framework & Ethical and ecological implications of digital biomarkers \\
\hline
Cogniscope & Social media engagement + semantic drift & Multimodal fusion improves robustness under noise. Achieves F1=0.58 for MCI and 0.92 for EarlyAD; explicitly models drift + noise for ecological validity. \\
\hline
\end{tabular}
\end{table}
\noindent \textbf{Simulation Baseline.} In addition to prior clinical benchmarks, we compare Cogniscope against its own simulation baseline (noise-free vs. noisy). Removing noise yielded accuracy of 0.947 (F1$_{MCI}$=0.868), while noisy conditions reduced accuracy to 0.850 (F1$_{MCI}$=0.582). This demonstrates how controlled noise injection increases ecological validity by reproducing diagnostic ambiguity observed in real-world MCI populations.

\subsection{Summary of Findings}

Three insights emerge:  
(1) \textit{Longitudinal semantic drift} captures decline more effectively than static coherence.  
(2) \textit{Behavioral entropy} is an early and robust marker, especially for MCI.  
(3) \textit{Multimodal fusion with noise injection} yields robust, ecologically valid classification.  

These findings support the use of computational modeling of engagement and language as scalable digital biomarkers of cognitive change. At the same time, they highlight MCI as the most challenging stage, requiring drift-aware, multimodal, and uncertainty-tolerant models.

\section{Conclusion}

In this work, we presented Cogniscope, a simulation framework that models social-media–style interactions as digital biomarkers of cognitive health. By formalizing behavioral and linguistic markers, calibrating simulation parameters against public engagement statistics, and conducting ablation and sensitivity analyses, we demonstrate how controlled synthetic environments can provide insights into early detection tasks. Beyond experimental results, we contribute an open-source benchmark testbed that enables reproducibility and invites the community to extend and validate our approach on real-world data. Ultimately, we position simulation as an ethically safe and practically valuable bridge toward future large-scale studies of cognition and online behavior.
\paragraph{Limitations and Future Work.} 
Cogniscope is a reproducible proof-of-concept built entirely on synthetic data, with engagement distributions benchmarked against public YouTube and TikTok statistics (Table~\ref{tab:validation}). The framework abstracts away from the full complexity of social media platforms, and the synthetic population lacks demographic and cultural diversity, limiting generalizability. Real-world validation remains essential. Future work includes testing on public datasets such as eRisk, ADReSSo, or Reddit health communities, incorporating demographic and cultural variability to improve fairness, scaling to larger populations and longer trajectories with temporal neural models, and exploring access to anonymized platform-level traces for more accurate behavioral calibration. Bridging simulation with empirical social traces will refine Cogniscope into a tool that demonstrates methodological novelty and provides actionable insights for health research and the ICWSM community.


\section*{Ethical Considerations}
This study uses only synthetic data generated through large language models and simulation, avoiding direct privacy risks. No personal or platform user data were collected or analyzed. However, we acknowledge that future validation on real-world traces (e.g., Reddit health communities, eRisk, or ADReSS datasets) will require careful attention to ethics and governance. Any such work must ensure informed consent where appropriate, comply with platform terms of service, and undergo institutional review board (IRB) oversight. In line with GDPR and HIPAA principles, personally identifiable information (PII) must never be stored, and derived signals should not be used for individual-level diagnosis without clinical validation. We also recognize the potential risk of stigmatization if cognitive predictions are misapplied. To mitigate this, we commit to framing Cogniscope strictly as a research tool for understanding online health signals, not as a diagnostic system. Ethical safeguards and transparency remain central to all planned extensions of this work.


\section*{Broader Impact}
Cogniscope contributes to an emerging line of research that uses computational social science methods to understand health and well-being through online behavior. By showing how social media–style engagement traces can serve as unobtrusive indicators of cognitive change, this work highlights the potential of digital platforms to support early intervention, especially in populations that may lack access to frequent clinical testing. At the same time, we recognize the ethical challenges of applying such methods in practice: any deployment must safeguard user privacy, ensure informed consent, and avoid stigmatization or misuse of health-related inferences. As noted in our \textit{Limitations} section, the current study relies on synthetic data, with engagement distributions loosely benchmarked against public statistics; future validation with real-world traces will require strict adherence to platform terms of service, institutional review board (IRB) oversight, and regulatory frameworks such as GDPR and HIPAA (see \textit{Ethical Considerations}). More broadly, this project underscores the dual responsibility of leveraging online data for societal benefit while maintaining respect for user autonomy and fairness across demographic groups.

\bibliography{aaai2026}
\end{document}